\documentclass[11pt]{article}

\usepackage{sectsty}
\usepackage{graphicx}
\usepackage{authblk}
\usepackage[utf8]{inputenc}
\usepackage[english]{babel}
\usepackage{url}

\usepackage{natbib}
\newcommand{\citeg}[1]{\cite[e.g.][]{#1}}

\title{
\textbf{Using Neural Networks to Learn the Jet Stream Forced Response from Natural Variability}}

\author[1]{Charlotte Connolly}
\author[1]{Elizabeth A. Barnes}
\author[2,3]{Pedram Hassanzadeh}
\author[4]{Mike Pritchard}
\affil[1]{Department of Atmospheric Science, Colorado State University, Fort Collins, CO}
\affil[2]{Department of Mechanical Engineering, Rice University, Houston 77005 TX}
\affil[3]{Department of Earth, Environmental and Planetary Sciences, Rice University, Houston 77005 TX}
\affil[4]{Department of Earth System Science, University of California, Irvine, CA 92697, USA}

\begin{document}
\maketitle

\begin{center}
This manuscript has been submitted for consideration for publication in \\
\emph{Artificial Intelligence for the Earth Systems}
\end{center}

\newpage

\begin{abstract}
Two distinct features of anthropogenic climate change, warming in the tropical upper troposphere and warming at the Arctic surface, have competing effects on the mid-latitude jet stream’s latitudinal position, often referred to as a “tug-of-war”. Studies that investigate the jet’s response to these thermal forcings show that it is sensitive to model type, season, initial atmospheric conditions, and the shape and magnitude of the forcing. Much of this past work focuses on studying a simulation’s response to external manipulation. In contrast, we explore the potential to train a convolutional neural network (CNN) on internal variability alone and then use it to examine possible nonlinear responses of the jet to tropospheric thermal forcing that more closely resemble anthropogenic climate change. Our approach leverages the idea behind the fluctuation-dissipation theorem, which relates the internal variability of a system to its forced response but so far has been only used to quantify linear responses. We train a CNN on data from a long control run of the CESM dry dynamical core and show that it is able to skillfully predict the nonlinear response of the jet to sustained external forcing. The trained CNN provides a quick method for exploring the jet stream sensitivity to a wide range of tropospheric temperature tendencies and, considering that this method can likely be applied to any model with a long control run, could lend itself useful for early stage experiment design.
\end{abstract}

\section{Introduction}\label{INTRODUCTION}
The eddy-driven jet stream drives much of the northern hemisphere mid-latitude weather \citeg{Nakamura2004, Athanasiadis2010, shaw2016, Madonna2017}. Consequently, changes in the jet stream position and strength can result in enormous societal impact by impacting heat waves, droughts, and flooding events \citep{Schubert2011, Coumou2012, Bibi2020, Rousi2021, Rousi2022}, extreme weather across the mid-latitudes \citep{Mahlstein2012, Rothlisberger2016}, hurricane tracks \citep{Mattingly2015}, and crop production \citep{Kornhuber2019}. Two robust features of anthropogenic climate change, warming in the upper troposphere of the tropics and warming at the surface of the Arctic, have been shown to independently force opposite responses in the mean jet location \citeg{Held1993,  Harvey2015, Stendel2021} These competing responses are driven by changes in the pole to equator temperature gradient \citep{Blackport2020, Stendel2021}. Warming in the tropical upper troposphere drives a poleward shift in the mean jet location by increasing the upper tropospheric temperature gradient, while simultaneously, warming at the Arctic surface drives an equatorward shift in the mean jet location by decreasing the surface temperature gradient \citep{Butler2010, Screen2013, Chen2020, Stendel2021}, The competing jet response stemming from these two thermal forcings is commonly referred to as the “tug-of-war” on the jet stream. Current consensus across climate models is that the upper tropospheric warming wins-out over the Arctic surface warming, causing a net poleward shift of the jet \citep{Yin2005, Swart2012, Barnes2013, Harvey2015}. However, there is still substantial disagreement over the magnitude of the jet response due to uncertainty in the strength and spatial extent of the regional heating anomalies \citep{Grise2016}.

Warming in both the tropical upper troposphere and Arctic surface are caused by distinctly different dynamical processes that determine the characteristics of the thermal anomalies. The tropical upper atmosphere warms more as a result of additional water vapor stored in the warmer tropical tropospheric air (i.e. a reduction in the moist adiabatic lapse rate; \citep{Sherwood_and_Nidhi_Nishant2015}. The enhanced Arctic warming, commonly referred to as Arctic Amplification, is occurring three times faster than elsewhere on the planet \citep{Blunden2012, Druckenmiller2021} and is driven by multiple processes that include changes in poleward energy transport \citep{Hwang2010,  Graversen2019}, surface ice-albedo feedbacks \citep{Manabe1980, Dai2019}, cloud feedbacks \citep{Abbot2008} and lapse-rate feedbacks \citep{Pithan2014}. To further increase the complexity of the processes driving the regional warming, the two thermal forcings likely do not act entirely independently. Research has shown that increased transient Rossby waves initiated in the tropics may drive increased heat transport into the high latitudes and as a result drive further warming in the mid- and upper-troposphere of the Arctic \citep{Lee2014, Dunn-Sigouin2021}. Uncertainties in the processes that contribute to the magnitude and shape of warming in the tropical upper troposphere and Arctic surface \citep{Blackport2020, Stendel2021}, in turn, make it even more challenging to predict the magnitude of the jet response. 

Despite the large body of work that investigates the response of the mid-latitude jet under climate change, multiple challenges, such as the short observational record, isolating the jet’s forced response from internal variability, and modeling ice and cloud feedbacks continue to make the question difficult to answer \citep{Tjernstrom2008, Kattsov2010, Cohen2014, Pithan2014, Vihma2014}. The studies that have investigated the response of the jet to a thermal forcing have shown that the jet is sensitive to the shape, location, and magnitude of the thermal forcing \citep{Butler2010}, the season in which the forcing is imposed \citep{McGraw2016}, the current state of the atmosphere (i.e. position of the jet stream; \citep{Gerber2008, Barnes2010, Kidston2010, Garfinkel2013}, and the climate models used for the study \citep{Meehl2007, Barnes2013}. 

In an attempt to explore circulation sensitivities to a wide range of possible thermal forcings, Hassanzadeh and Kuang (2016b) used a control run from the GFDL dry dynamical core \citep{Manabe1974} and employed the fluctuation-dissipation theorem (FDT) to compute the linear response function of the circulation to a number of external thermal and mechanical forcing. FDT relates the mean linear response of a nonlinear system to a forcing through a linear operator created from the internal variability of the system \citeg{Kraichnan1959, Leith1975, Marconi2008}. With the ability to explore a forced response from internal variability, FDT has been proposed as a method to quickly estimate circulation sensitivities in climate models \citep{Fuchs2015} and serve as a useful tool for planning expensive climate model experiments \citep{Leith1975}. There have been encouraging results using FDT to explore the circulation response to thermal forcings in general circulation models \citep{Gritsun2007} as well as more complex coupled climate models \citep{Phipps_undated} to estimate the response to realistic sea surface thermal forcings \citep{Fuchs2015}. 

In order for the linear operator of FDT to accurately predict the mean response to a forcing, the system must satisfy a number of conditions \citep{Marconi2008}. The first condition is that the system must be in equilibrium, because FDT assumes that small changes in the system’s state (internal variability) has a recovery back to equilibrium that is similar to the system’s response to a small perturbation  \citep{Kraichnan1959, Leith1975}. The second is that the perturbation must be small enough so that the response is linear even though the system that the operator is created from is not necessarily linear (Leith 1975). Lastly, the probability density function of the system must be differentiable, and many applications of the FDT assume the system probability density function is Gaussian \citep{Majda2005}, though work has been done to make versions of FDT where the system can be quasi-Gaussian \citep{Cooper2011}. In theory, a system that satisfies these conditions can use FDT to compute the systems’ linear response to a forcing, though there are practical challenges in applying FDT to high-dimensional systems, such as GCMs \citep{Lutsko2015, Hassanzadeh2016_p2, Khodkar2018}.

Instead of using FDT to relate a forcing to a response, this study uses a convolutional neural network (CNN) to learn the nonlinear relationship between a forcing and a response. Moreover, using a CNN in place of the linear operator removes the need to make some of the FDT assumptions (i.e. small forcing for a linear response, Gaussianity assumption). Training is performed on data from a long control run with the CESM dry dynamical core. Once trained, the CNN is used to explore the jet sensitivity to a variety of thermal forcings. Throughout this study, we evaluate the CNN’s ability to quantify the CESM dry dynamical core's jet sensitivity, placing particular emphasis on the tug-of-war between the warming in the tropical upper troposphere and the Arctic surface. Training a neural network on internal variability alone and then using it to predict a forced response is, to our knowledge, a novel application of deep learning to climate analysis. Therefore, we assess the strengths and weaknesses of this approach in multiple ways (see Section \ref{RESULTS}). 

\section{Methods}\label{METHODS}

We train a CNN to predict the jet stream response to zonally averaged regional temperature perturbations. The goal is to investigate jet sensitivity to thermal features associated with anthropogenic climate change. The CNN, detailed in Section \ref{CNNS}, is trained on a long control run from a dry dynamical core, which is documented to reproduce the majority of the northern hemisphere’s jet response to heating perturbations along with simulating the correct sign of the jet shift \citeg{Mbengue2013, Hassanzadeh2014, McGraw2016, Baker2017}. Once trained, the CNN’s skill is examined by comparing to additional baseline prediction methods and dry core experiments that include an imposed thermal forcing. Details on the dry dynamical core setup, the CNN architecture and training, additional baseline prediction methods, and additional dry core heating experiments are discussed in more detail in the following Sections.

\subsection{Training Data}\label{TD}

We use output from the Community Earth System Model (CESM) Eulerian spectral-transform dry dynamical core \citep{Lauritzen2018}. The model runs are completed with the Held-Suarez configuration \citep{Held1994}, such that friction exists at the surface and the temperature is relaxed to a prescribed hemispherically symmetric temperature field. The relaxation temperature field is set to equinoctial conditions and there is no absorption of solar energy by the atmosphere (i.e. there are no seasons or diurnal cycles). All runs are performed at T42 resolution with 30 vertical levels, 64 latitude bands, and 128 longitude bands. The simulation is run in the above configuration for one million six hour time steps. The first 20,000 time steps (13.7 years) are thrown out to account for model spin up.

All data processing is performed to create efficient training data for a CNN (Section \ref{CNNS}) to predict the zonally-averaged jet response to a range of tropospheric temperature perturbations. Two variables are used in this study, zonally averaged temperature and zonally averaged zonal wind speed. The zonally-averaged temperature data is used to calculate the temperature tendency field used as input to the CNN. The zonally-averaged zonal wind speed is used to calculate the initial location of the jet and the subsequent shift of the jet, which are used as a CNN input and the CNN prediction respectively. We exclude data from 200 hPa and above, effectively removing the stratosphere which is not well resolved in this model without modification \citep{Polvani2002} and so focus can remain solely on the troposphere for both the forcing and the jet response. Given that this study focuses on hemispheric jets, we take advantage of the hemispheric symmetry in the dry core and use each hemisphere as a separate independent sample, doubling the amount of available training data to two million. After zonally averaging, removing the stratosphere, and considering each hemisphere as a separate sample, the resulting size of the temperature field is 25 vertical levels by 32 latitude bands.

Backward differencing is used to calculate the temperature tendency which is then smoothed using a backward running mean of 240 time steps (60 days) to remove higher frequency variability. Removing the high frequency variability allows the network to focus on learning the response to a forcing that more closely mimics continuous climate change forcing. This smoothing is also aligned with FDT calculations, in which an integration over long time lags (often up to the decorrelation timescale) is done; e.g., see Eq. (3) in Hassanzadeh and Kuang (2016b). Smoothing the data before calculating the temperature tendency did not result in any changes in the CNN skill (not shown).

Following established methods \citep{Woollings2010, McGraw2016}, the jet location is defined as the latitude of the maximum wind speed at a pressure level near the surface. Zonal wind speeds from the 850 hPa level are used here and are first smoothed with a 240 time step (60 days) backward running mean. Then, a second order polynomial is fit to the peak of the smoothed 850 hPa zonal wind profile and the jet location is defined as the latitude of the maximum wind speeds. 

Now that a smoothed zonal temperature tendency and a jet location are calculated, the data are split into training, validation, and testing data. Splitting is completed by chunking the data into three groups where training data is the first chunk, validation the second, and testing the last. 

Lastly, the jet response to a given temperature tendency is defined by the change in jet latitude from the time of input to 120 time steps later (i.e. the jet shift). A positive jet shift indicates a poleward shift in jet location and a negative jet shift indicates an equatorward shift in jet location relative to the jet’s latitude at the time of prediction. The jet shift is calculated within each dataset (training, validation, and testing) by subtracting a 240 time step backward running mean of jet stream locations from a 240 time step forward running mean of jet stream locations 120 time steps into the future. This processing results in 359,280 training samples, 199,280 validation samples, and 1,399,558 testing samples. Training the CNN required fewer samples than expected as adding more samples to the training dataset did not improve the CNN skill, explaining why the testing dataset is much larger than the training and validation datasets.

\subsection{Convolutional Neural Network}\label{CNNS}

\begin{figure}[h]
 \centerline{\includegraphics[width=40pc]{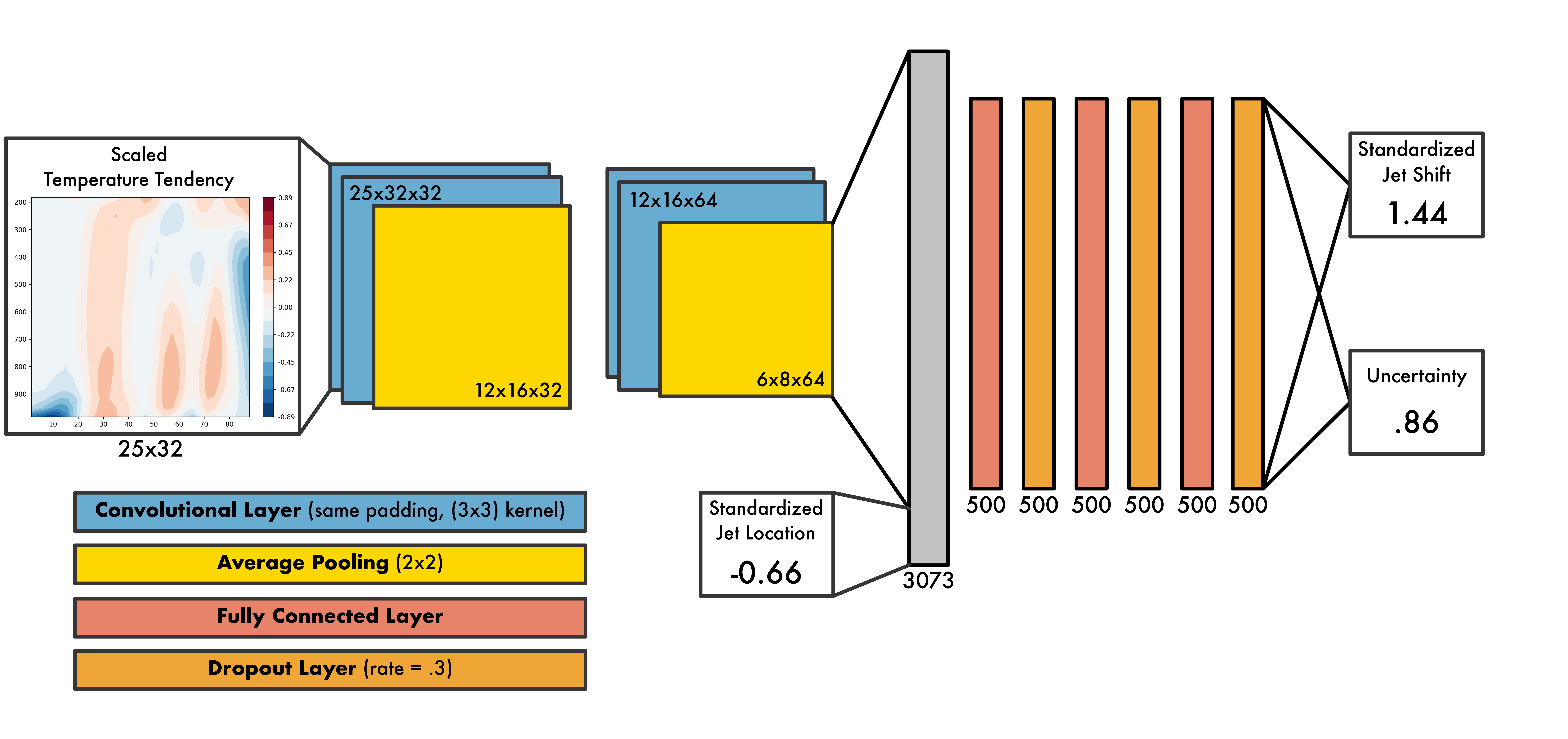}}
  \caption{Schematic of the convolutional neural network with an example of an input and output.}\label{arch}
\end{figure}

CNNs are commonly used for image recognition and classification tasks as the convolutional layers can extract spatial features in the input image that help the network learn the correlations between the inputs and output \citep{Fukushima1980, YannLeCun1998, Zeiler2014}. While a fully connected feedforward network \citeg{LeCun2015} has the ability to learn the same features extracted by the convolutional layers within a CNN, it may require a larger network and more training data to do so \citep{YannLeCun1998, Ingrosso2022}. In this study, we utilize a CNN so that the network can efficiently learn the correlation between temperature tendencies and the jet response while also trying to minimize the amount of training data required.

The CNN has two inputs: a smoothed temperature tendency field (K day$^{-1}$) and an initial jet location (degrees latitude). Including the temperature tendency as an input allows us to investigate the jet response to regional temperature tendencies and including the initial jet location supplies the CNN with essential information about the current state of the jet at the prediction time, an important factor for the jet response to forcing \citep{Gerber2008, Barnes2010, Kidston2010, Garfinkel2013}. Before the data is input into the CNN, the smoothed temperature tendency field is multiplied by a factor of 10 and the initial jet location is standardized using the standard deviation and mean jet location from the training data. Scaling and standardizing are done so that both inputs have similar magnitudes (order of 1). 

The network consists of four convolutional layers: two average pooling layers, three dense layers, and three dropout layers (Fig. \ref{arch}). Convolutional and dense layers use the hyperbolic tangent activation function. Data are passed through the network as follows: the scaled temperature tendency goes directly into the first convolutional layer with 32 filters of size 3 x 3 and a stride of 1 followed by a second convolutional layer with the same attributes. The second convolutional layer is then connected to an average pooling layer with a kernel size of 2 x 2. These three layers, two convolutional and a single average pooling, are repeated with the same attributes with the exception of containing 64 filters rather than 32 in the convolutional layers. The output from the second average pooling layer is flattened and the standardized initial jet location is concatenated to the end. This layer is then fed into the first dense layer with 500 nodes and then goes through a dropout layer with a dropout rate of 30\%. The data passes through a combination of dense layers with 500 nodes followed by dropout layers with a dropout rate of 30\% two more times. The data from the final dropout layer then passes into the output layer consisting of two nodes.

The CNN outputs two values denoted as $\mu$ and $\sigma$, which represent a mean and standard deviation of a Gaussian distribution where $\mu$ denotes the predicted jet shift and $\sigma$ represents its uncertainty. Predicting the parameters of a Gaussian distribution is commonly used to quantify uncertainty for neural networks \citep{Nix1994, Nix1995}, and \citet{Gordon2022} recently showed the utility of incorporating uncertainty into a regression neural network for climate science applications. The network learns to predict $\mu$ and $\sigma$ through the implementation of the negative log-likelihood loss function: 

\begin{equation}
L_{i} =  -log(p_{i})
\label{loss function}
\end{equation}

where $p$ is a value of the predicted Gaussian distribution evaluated at the true jet shift for the $i^{th}$ input sample. To ensure the network is calibrated we employ the probability integral transform (PIT) probability calibration scheme \citep{Gneiting2007, Nipen2011, Barnes2022}. The PIT histogram for this CNN can be found in Supplementary.

To train the CNN, we use the Adam stochastic gradient descent optimization algorithm \citep{Kingma2014} with a learning rate of $10^{-7}$, a batch size of 256, and a random seed of 300. We apply early stopping to halt the training process once the validation loss fails to decrease for 10 consecutive epochs and restore the model weights to the version with the lowest validation loss \citep{Prechelt2012}. 

\subsection{Baselines}\label{BASELINES}

We establish two baselines in this study to assess the performance of the CNN and demonstrate that the CNN has learned relationships between the jet response and the regional temperature tendencies. The first baseline is called \emph{persistence}, similar to “persistence forecasting” \citep{Mac_Donald1992}, where future conditions are predicted to be identical to the current conditions. In our case, this translates to the jet’s future location being the same as its location at the time of prediction (i.e. jet shift equal to zero). Comparing this baseline to the CNN ensures that the CNN is predicting jet shifts that are more accurate than predicting a jet shift of zero. The second baseline is called \emph{average evolution} and describes the average movement of the jet based on its position at the time of prediction. For this baseline calculation, the training data is separated into 100 different bins according to the initial jet location, essentially grouping samples with similar initial jet locations together. The average jet shift for each bin is calculated, resulting in an average jet response that is solely dependent on the jet stream’s starting position. The average evolution baseline is not sensitive to the number of bins or their exact spacing (not shown). This baseline ensures that the CNN is not just predicting the average evolution of the jet based solely on the initial jet location but is also using the temperature tendency input to make its prediction.

Every test sample is thus associated with three jet shift predictions, one from the CNN and two from the additional persistence and average evolution baselines. Although comparing results between the CNN and the two baselines is useful for placing the CNN’s predictions into context, we highlight that the baselines make predictions based solely on information about the initial location of the jet while the CNN is provided additional information in the form of the temperature tendency. Thus, the CNN is able to explore the correlations between a temperature tendency and a jet response.

\subsection{Heating Experiments}\label{HE}

The main goal of this study is to investigate the jet stream sensitivity to thermal forcing driven by anthropogenic climate warming. However, as we have designed it, the CNN only trains on data from a long control run (i.e. internal variability), and thus, only provides insights into the forced response if the idea of the FDT holds \citep{Kraichnan1959, Leith1975, Marconi2008}. To investigate whether this assumption is valid, we run additional dry core simulations (referred to as \emph{heating experiments}) with zonally symmetric imposed thermal forcing  (F) that take the form of a two-dimensional Gaussian in the latitude/pressure plane (Equation \ref{guassian blab}):

\begin{equation}
F(\Theta,p) = q_o exp \left[ \frac{(\mid \Theta \mid - \Theta_o)^2}{\Theta^2_w} - \frac{(\mid p \mid - p_o)^2}{p^2_w} \right]
\label{guassian blab}
\end{equation}
\\

where $\Theta_o$ and $p_o$ are the horizontal and vertical centers respectively, $\Theta_w$ and $p_w$ define the width and height, and the magnitude of the forcing is given by $q_o$. Gaussians that fall near the edge are cut off and therefore not complete two-dimensional Gaussians. For all Gaussians created in this study, $\Theta, \Theta_w, p, p_w, q_o$ are reported in Appendix A. 

Eighteen heating experiments with either one or two Gaussian thermal forcings imposed are run out to equilibrium to quantify the true jet shift (see experiments \#1-18 in Table A1). To perform a direct comparison between the CNN-predicted jet responses and the dry core jet responses, the CNN is given the same temperature tendency that is imposed in each of the dry core heating experiments. For the CNN’s initial jet location input, the average jet location from the long control run is used (42.4\textdegree). By comparing the true forced response from the dry core to the predicted forced response by the CNN we are able to investigate the CNN's ability to predict the jet response to thermal forcing from training on internal variability alone.

Each heating experiment is initiated at the end of the long control run (time step one million; 684.9 years), and therefore, have the same initial conditions. The heating experiments are run for an additional 20,000 time steps with the first 4,000 removed to ensure that the model has reached its new equilibrium. The 850 hPa zonal winds are then used to compute the location of the jet (see Section 3.1). Finally, the true response of the jet to an imposed thermal forcing is defined as the average jet location during the long control run subtracted from the jet location in the corresponding heating experiment. 

\section{Results}\label{RESULTS}

\subsection{Evaluation of CNN skill}\label{ECNN}

\begin{figure}[h]
 \centerline{\includegraphics[width=40pc]{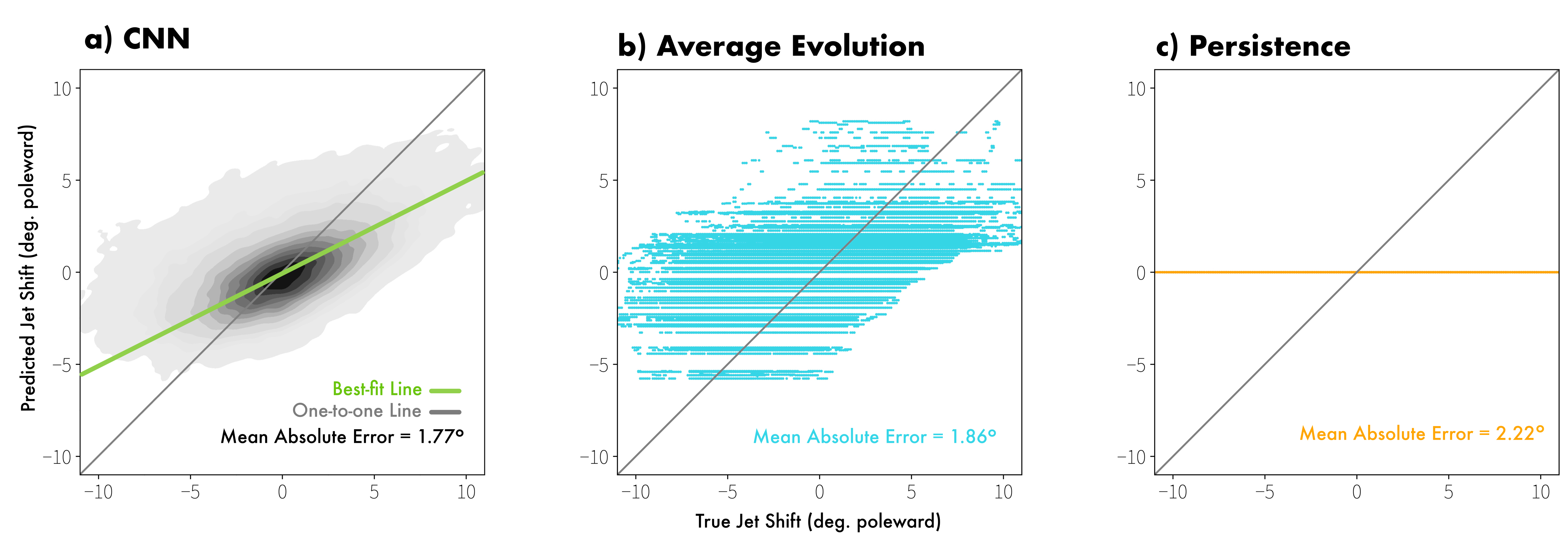}}
  \caption{Predicted jet response (y-axis) versus the true jet response (x-axis) for the (a) CNN, (b) average evolution baseline, and (c) persistence baseline using the testing data from the control simulation. Panel (a) is a contoured by density and panels (b) and (c) are scatter plots. Mean absolute errors are shown in the bottom right corner of each panel. Gray lines represent a perfect prediction (one-to-one line). Green line in panel (a) represents the best-fit line from the CNN predictions.}\label{scatter}
\end{figure}

We begin our discussion of the results with a focus on the deterministic predictions by the CNN ($\mu$). The deterministic skill on the testing data, which we define as the mean absolute error between the predicted jet shift and the true jet shift, reveals how well the CNN generalizes to unseen samples within the control simulation. The first look at the entire testing dataset will appear to show a modest difference; a closer look within the testing dataset will prove more interesting. Figure \ref{scatter}a shows the relationship between the predicted jet shift and the true jet shift where predictions with higher accuracy are closer to the gray diagonal line (one-to-one line). Using orthogonal distance regression \citep{Boggs1990, SciPy2020}, which takes into account error in both the x and y variables as well as the CNN-predicted uncertainties in y, we calculate the slope from the testing data to be 0.5 deg. poleward / (deg. poleward). This positive slope demonstrates the CNN has learned relationships between the jet shift and the inputs. However, the slope of the CNN predictions is less than that of the one-to-one line implying that the CNN underestimates the magnitude of the largest jet shifts. This is likely a result of the imbalanced training data as it includes more samples with smaller jet shifts than larger ones (shown in Figure \ref{scatter}a by the density contours). During training, the goal of the CNN is to minimize the negative log-likelihood loss function (see Equation \ref{loss function}), but with an unbalanced dataset, the CNN may never predict the most extreme cases. Applying methods to make the network predict extremes, such as balancing the dataset, using samples weights, or creating custom loss functions \citep{Ma2013, Krawczyk2016}, either caused a severe decrease in skill or did not succeed in solving the problem (not shown). Nonetheless, as we will show next, the CNN outperforms the two benchmark baselines and is an effective tool for exploring jet sensitivity to external forcing.

Comparing the CNN’s skill on the testing data to that of the two baselines: average evolution (Fig. \ref{scatter}b) and persistence (Fig. \ref{scatter}c), allows us to place the CNN’s skill into context against other basic prediction methods. Persistence has the lowest performance with a mean absolute error of 2.22\textdegree. Average evolution performs only slightly worse than the CNN with mean absolute errors of 1.86\textdegree\ and 1.77\textdegree\ respectively. Unlike persistence, which can only ever predict a jet shift of zero, average evolution makes a prediction based on the average relationship between the initial jet location and the jet shift of the training data, allowing it to capture the mean jet response. Regardless, average evolution is limited to predicting one of the 100 jet shifts resulting from the methods used to calculate it (see Methods), hence the stripes in Figure \ref{scatter}b. Based on the mean absolute error alone, the CNN outperforms both the persistence and average evolution baselines for the testing data from dry dynamical core long control run. 

\begin{figure}[h]
 \centerline{\includegraphics[width=40pc]{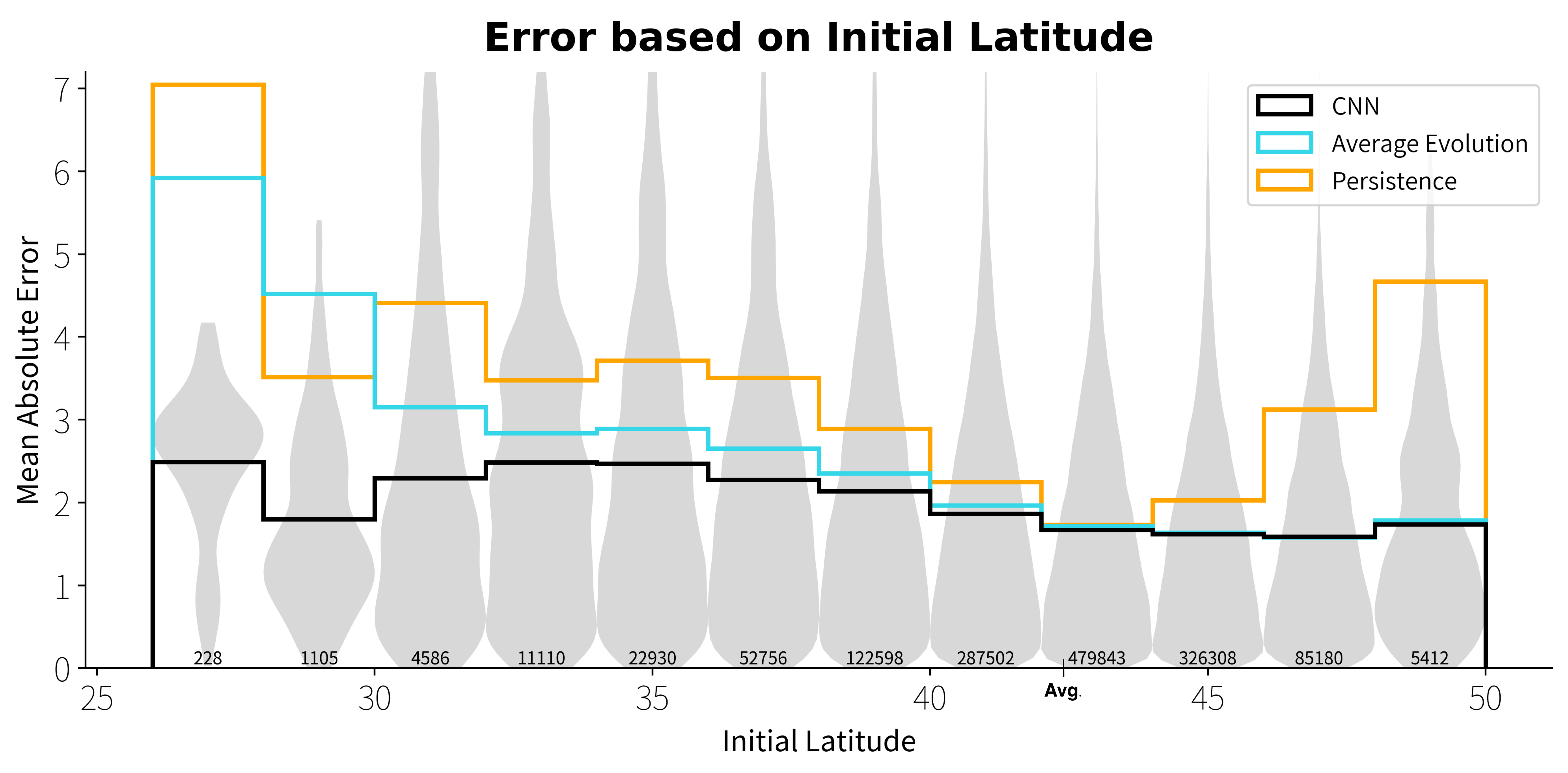}}
  \caption{The mean absolute error from the three prediction methods, CNN (black line), average evolution (cyan line), and persistence (orange line) grouped by initial jet locations. gray violin plots show the density curves of the CNN’s error distribution where the width corresponds with the frequency of the data. Numbers at the bottom of each bar indicate the number of samples in each group and the average initial jet location from the training data is marked on the x-axis (42.4\textdegree).}\label{violin}
\end{figure}

The mean absolute errors in Figure \ref{scatter} represent the error over the entire testing set, which is prone to obscuring interesting details hiding within the distribution. For a more comprehensive analysis of the CNN’s skill, the testing data is thus separated into groups based on the initial jet location. The mean absolute error for each group is shown for the CNN and the baselines in Figure \ref{violin} and describes how the CNN’s skill and the baselines' skill depend on the initial state of the jet. The gray violin plots behind each bar indicate the CNN’s mean absolute error distribution within that bin (i.e. the data used to calculate the CNN mean absolute error in each group). The violin plots are smoothed with a kernel estimator using Scott’s rule and 100 points, which are the default parameters of the Matplotlib library \citep{Hunter2007}. The numbers at the bottom of each bar denote the number of samples in that bin. For all bins in Figure \ref{violin}, the CNN outperforms the baselines as demonstrated by the CNN’s error (black line) falling below the baseline errors (cyan and orange lines). When the initial jet location is equatorward of 42\textdegree\ (labeled as “Avg.” along the x-axis of Fig. \ref{violin}), the CNN does considerably better than the baselines, but when the jet location is poleward of 42\textdegree, the CNN and average evolution achieve similar skill. That is, in the cases where the initial jet is near the pole, it appears that the CNN does not learn more than average evolution but instead learns this average behavior to make its prediction. 

When the initial jet location is near this climatological average position (42.4\textdegree), the errors of the CNN and baselines converge (Fig. \ref{violin}). About 30\% of the samples in the training data have initial jet locations within 2\textdegree\ of the climatological average and 14\% of these samples have a jet shift between -0.5\textdegree\  and 0.5\textdegree. Since so many samples near the climatological average have small jet shifts and because the persistence baseline can only predict a jet shift of zero, the mean absolute error for persistence is at its lowest near the jet’s climatological average position. The average evolution baseline converges to a near zero prediction when the initial jet location is near the climatological average, resulting in persistence and average evolution exhibiting similar errors. Although the persistence and average evolution baselines have an advantage near the climatological average, the CNN still outperforms both baselines, implying that the CNN is using the additional information provided by the input temperature tendencies.

\begin{figure}[h]
 \centerline{\includegraphics[width=45pc]{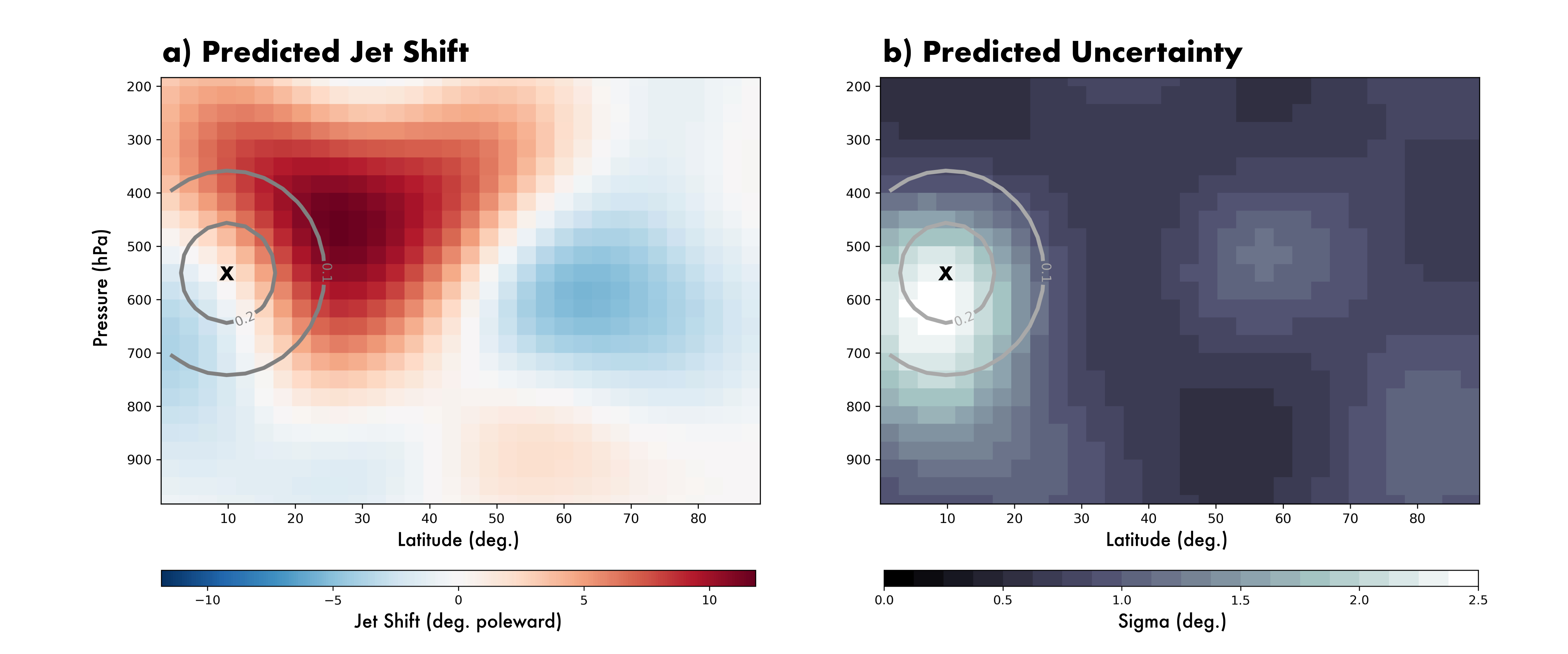}}
  \caption{The thermal forcing with a magnitude of q$_o$ = 0.25 K day $^{-1}$ is moved around the latitude and pressure plane where the shading in (a) and (b) represent the CNN-predicted jet shift and uncertainty respectively. An example of a thermal forcing is seen in the gray contours in both panels. The “x” marks the center as well as the predicted jet shift and predicted uncertainty associated with that thermal forcing. Gaussian parameters are found in Table A1 experiment \#19.}\label{blab}
\end{figure}

Next, we focus on evaluating CNN’s ability to predict a jet stream forced response from an artificially constructed idealized temperature tendency not encountered within its noisy training environment. These temperature tendency inputs contain a two-dimensional Gaussian (see Methods) with a prescribed magnitude, size, and location (latitude and pressure). Outside of the Gaussian, the temperature tendency field is filled with zeros. Although some of these thermal forcings have magnitudes larger than any temperature tendencies found in the internal variability training data, we will provide strong evidence to support the CNN’s ability to extrapolate in the coming sections. CNN predictions made from a thermal forcing use an initial jet location defined by the average jet location of the training data (42.4\textdegree). Therefore, differences in predicted jet shifts between temperature tendency inputs are a response to the thermal forcing alone and not the presumed initial state.

The shading in Figure \ref{blab} shows the CNN-learned jet sensitivity to the location of heating by holding the magnitude and the shape of a thermal forcing constant and changing only its location (Fig. \ref{blab}; see experiment \#19 in Table A1). An example of a thermal forcing is shown in the gray contours where the “x” denotes its center and the color of the shading beneath represents the predicted jet shift (Fig. \ref{blab}a) and the predicted uncertainty (Fig. \ref{blab}b) from the temperature tendency. 

Figure \ref{blab}a exhibits multiple known features of the jet response to tropospheric thermal forcings. For example, thermal forcings located higher in the troposphere are known to be more effective at perturbing the jet than thermal forcings located lower in the troposphere \citep{Hassanzadeh2016, Kim2021}. This feature is learned by the CNN and is shown in Figure \ref{blab}a as darker shading at higher pressure levels. In addition, warming in the tropical upper troposphere has been previously shown to cause the jet to shift poleward \citep{Chen2008, Lim2009, Butler2010}. This poleward jet shift is seen in Figure \ref{blab}a as denoted by the red shading in the tropical upper troposphere. Finally, heating at the polar surface has been shown to cause the jet to shift equatorward \citep{Butler2010, Deser2010, Screen2013} and this is seen in Figure \ref{blab}a by the blue shading, albeit weak, near the polar surface. These features indicate that the CNN has learned the correct sign of the jet shift as supported by prior research.

Figure \ref{blab}a also highlights how moving the center of the heating by a few degrees or pressure levels can change the direction of the jet shift. Take for instance heating at the polar surface, where moving the heating from 80\textdegree\  latitude to 75\textdegree\ latitude changes the jet response from an equatorward shift to a poleward shift. Baker et al. (2017) investigates the jet sensitivity to the location of heating by running 306 dry core experiments with an imposed Gaussian shaped temperature tendency that is moved around the latitude-pressure plane, just as we have done here with a trained CNN. In Baker et al. (2017), they show that changes in the latitude of the heating most strongly impact the sign of the jet shift while changing the pressure level has very little impact. Similar behavior is found here with the CNN, although with a few exceptions. A more in-depth discussion about the failure of the CNN to capture the correct direction of the jet shift at the surface of the mid-latitudes is found in Supplementary.

Recall that the CNN predicts both the jet shift ($\mu$) as well as its uncertainty ($\sigma$). Figure \ref{blab}b displays the predicted uncertainty values and highlights three regions where the CNN is less certain. The CNN is less certain when heating occurs around 10\textdegree\ latitude and 600 hPa ($\sigma \approx$ 4\textdegree) and additionally has large uncertainty when the heating is centered near 85\textdegree\ latitude and 900 hPa and the 60\textdegree\ latitude and 550 hPa ($\sigma \approx$ 1.5\textdegree). The reasons behind the greater uncertainty in these regions require further investigation.  

\subsection{Nonlinearities learned by the CNN}\label{NCNN}

Ideally, a benefit of using a CNN is that it learns a nonlinear relationship between the temperature tendency input and the jet shift output. The hyperbolic tangent activation functions in the convolutional and dense layers of the CNN allow it to learn nonlinear relationships between the inputs and outputs if nonlinearity is present in the data. However, this does not necessarily mean the CNN has learned nonlinear relationships. To evaluate the nonlinearity learned by the CNN we complete two analyses. The first analysis examines how the CNN-predicted jet shift varies as a function of the thermal forcing magnitude. The second analysis looks at scenarios where two thermal forcings are simultaneously present in the temperature tendency input and explores whether the CNN has learned a nonlinear interaction between the two. Keep in mind that neither of these analyses has a ground truth, and so we are exclusively exploring what the CNN has learned. In the next section, we will then further test the accuracy of the CNN with additional dry core simulations.

\begin{figure}[h]
 \centerline{\includegraphics[width=40pc]{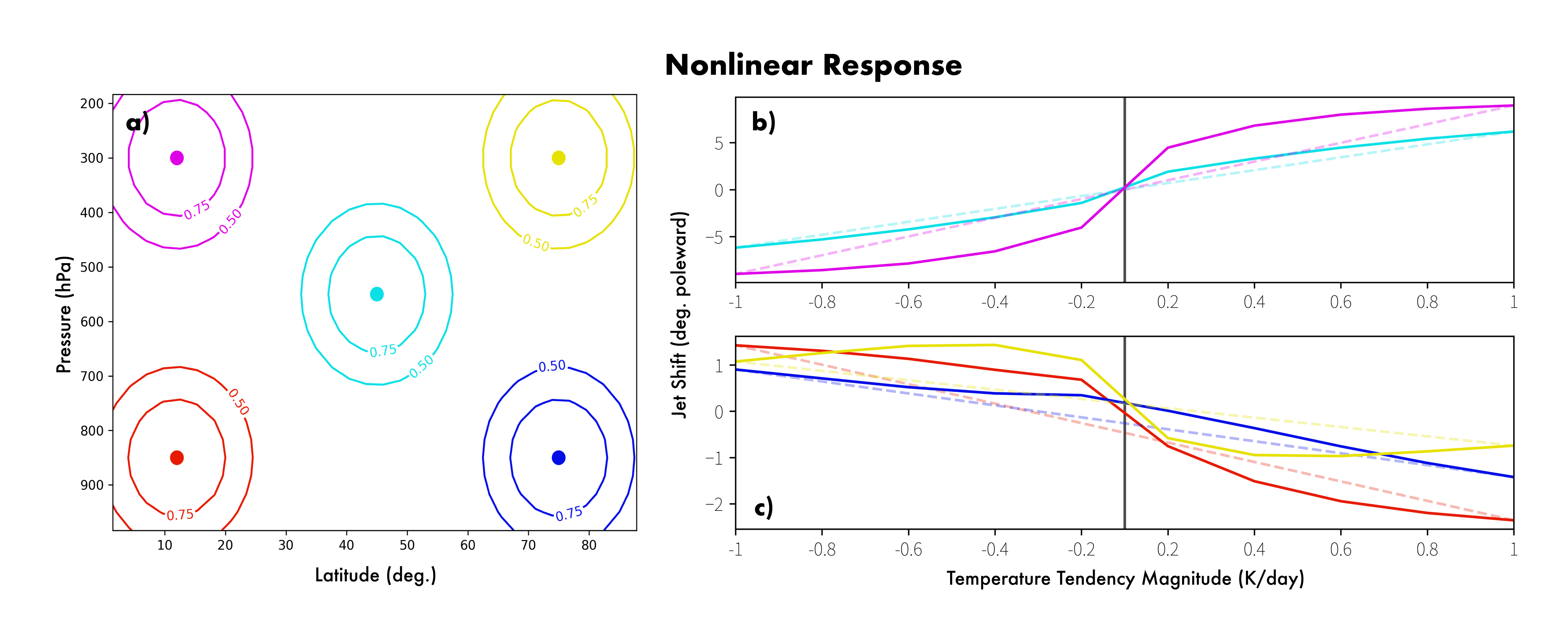}}
  \caption{The CNN learned nonlinear responses where panel (a) shows example thermal forcing with magnitudes of 1 K day$^{-1}$ at the five locations and panels (b) and (c) show the linear response (dashed line) and the predicted response from the CNN (solid lines). Note the different y-axes in panels (b) and (c). Gaussian parameters are found in Table A1 \#20-24.}\label{blablinear}
\end{figure}

The first nonlinear analysis explores how the jet shift varies as a function of the thermal forcing magnitude by separately inputting thermal forcings of different magnitudes in five different locations (Fig. \ref{blablinear}a; see experiments \#20-24 in Table A1). We use ten different magnitudes that vary from -1.0 K day$^{-1}$ to 1.0 K day$^{-1}$ in increments of 0.2 K day$^{-1}$ for each location. For all cases, the initial jet location input is fixed at the average jet location of the training data (42.4\textdegree). Figure \ref{blablinear}b and \ref{blablinear}c compare a linear relationship (dashed lines) and the CNN’s learned relationship (solid lines) between the thermal forcing magnitude and the jet shift. The jet response to temperature tendencies near the polar surface and the mid-tropospheric mid-latitudes are the most linear as shown by the green line in Figure \ref{blablinear}b and the blue line in Figure \ref{blablinear}c. In both of these cases, the CNN-predicted jet shift is most similar to the linear dashed line. In contrast, temperature tendencies in the tropics and the upper troposphere of the polar region have the largest nonlinear response (yellow line in Figure \ref{blablinear}b, pink, and red lines in Figure \ref{blablinear}c) as these cases vary greatly from the dashed linear line.

\begin{figure}[h]
 \centerline{\includegraphics[width=40pc]{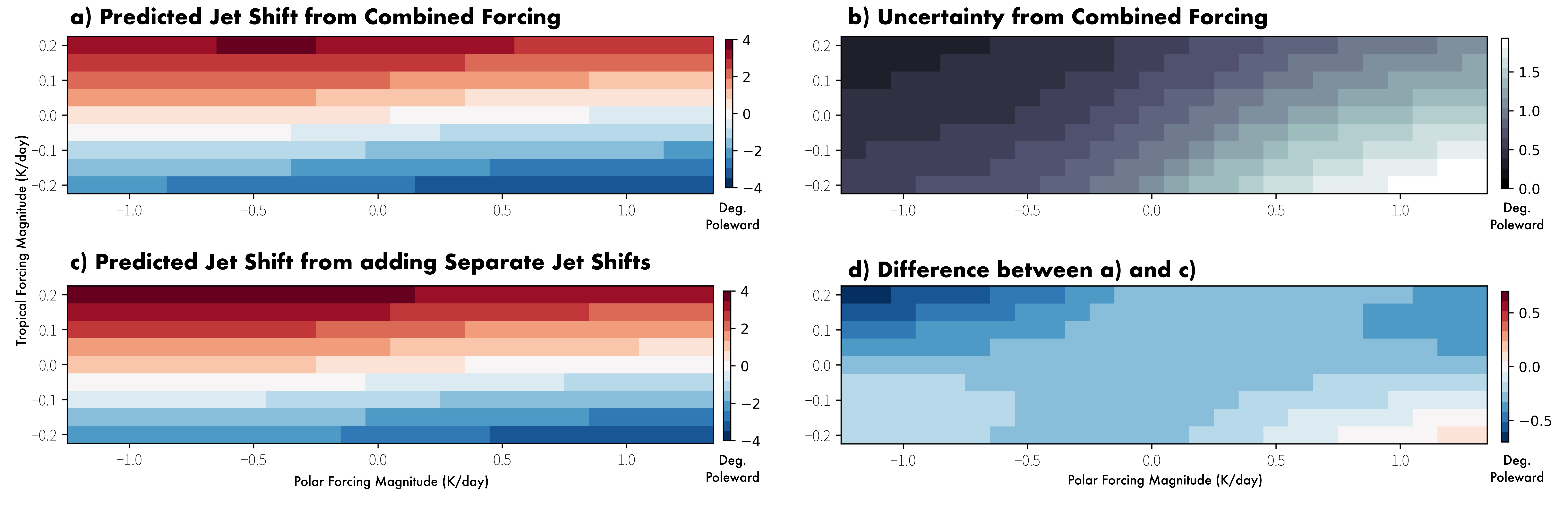}}
  \caption{Panel (a) shows the CNN-predicted jet shift from when two thermal forcings vary with magnitude. One at the surface of the pole (x-axis) and the other in the upper troposphere of the tropics (y-axis). Panel (b) shows the predicted uncertainty for the predictions of panel (a). Panel (c) is similar to panel (a) but the CNN predicts the jet shift from the two thermal forcings independently. Panel (d) is the difference between panel (a) minus panel (c) representing the CNN learned nonlinearity.}\label{war_LINEAR}
\end{figure}

We next explore the nonlinearities learned by the CNN when two thermal forcings are present. The thermal forcings are centered on two key regions, the tropical upper troposphere and the polar surface. As discussed previously, warming in the tropical upper troposphere forces the jet to shift poleward \citeg{Chen2008, Lim2009, Butler2010} and warming at the polar surface forces the jet to shift equatorward \citeg{Butler2010, Deser2010, Screen2013}. When they occur simultaneously, they force competing effects that can result in a tug-of-war scenario on the jet stream \citeg{Harvey2015, Chen2020}. To explore the jet sensitivities to this climate change induced tug-of-war, the temperature tendency inputs are composed of a Gaussian thermal forcing at the polar surface and another in the upper troposphere of the tropics. Both vary independently in magnitude during the analyses (see experiment \#25 in Table A1) and Figure \ref{war_LINEAR}a shows the predicted jet shift ($\mu$) and \ref{war_LINEAR}b shows the predicted uncertainty ($\sigma$) for each forcing pattern. With regards to the tug-of-war, studies use a variety of atmospheric models to show that despite opposite forced jet responses, the jet will likely shift poleward \citep{Yin2005, Harvey2015}. The upper right quadrant of Figure \ref{war_LINEAR}a depicts the situation where both thermal forcings are positive (warming). In this scenario, the CNN predicts a poleward shift of the jet in agreement with past work, however, the CNN is not equally certain for all predictions. As shown in Figure \ref{war_LINEAR}b, the CNN is more confident with cooling at the pole and warming in the tropics and less confident with warming in the pole combined with cooling in the tropics. Understanding why these scenarios are more uncertain requires further investigation.

Figure \ref{war_LINEAR}a shows the CNN-predicted jet response when two thermal forcings are present in the temperature tendency input. To test whether the CNN has learned a nonlinear impact on the jet from two simultaneous forcings, we task the CNN to predict the jet shift from the two thermal forcings independently (upper tropical troposphere and polar surface) and add the two predicted jet shifts together subsequently. If the CNN exclusively learned a linear response between two forcings, Figure \ref{war_LINEAR}a and Figure \ref{war_LINEAR}c would be identical, as predicting a jet shift from combined forcings would be equal to predicting the jet shifts from individual forcing and adding the predictions together. Instead, Figure \ref{war_LINEAR}d shows the difference in predicted jet shifts from these methods and provides evidence of the nonlinearity learned by the CNN where inputs that contain stronger thermal forcings (scenarios in the corners of  \ref{war_LINEAR}d) have greater learned nonlinearity.

\subsection{Out-of-sample tests}\label{OCNN}

Thus far we have compared the CNN-predicted jet shifts to our established baselines, true jet shifts harvested from the internal variability of the control run, and past work. We next evaluate the ability of the CNN to predict the explicit simulated jet response to an imposed idealized steady thermal forcing outside the training set. The FDT states that the linear response of a nonlinear system to external forcing can be related to the internal variability of the system. Under the assumption that FDT holds, our CNN trained on internal variability may also be able to predict a forced response. We next explore this by comparing the true forced jet shift calculated from additional dry core experiments to the predicted jet shift by the CNN.

\begin{figure}[h]
 \centerline{\includegraphics[width=30pc]{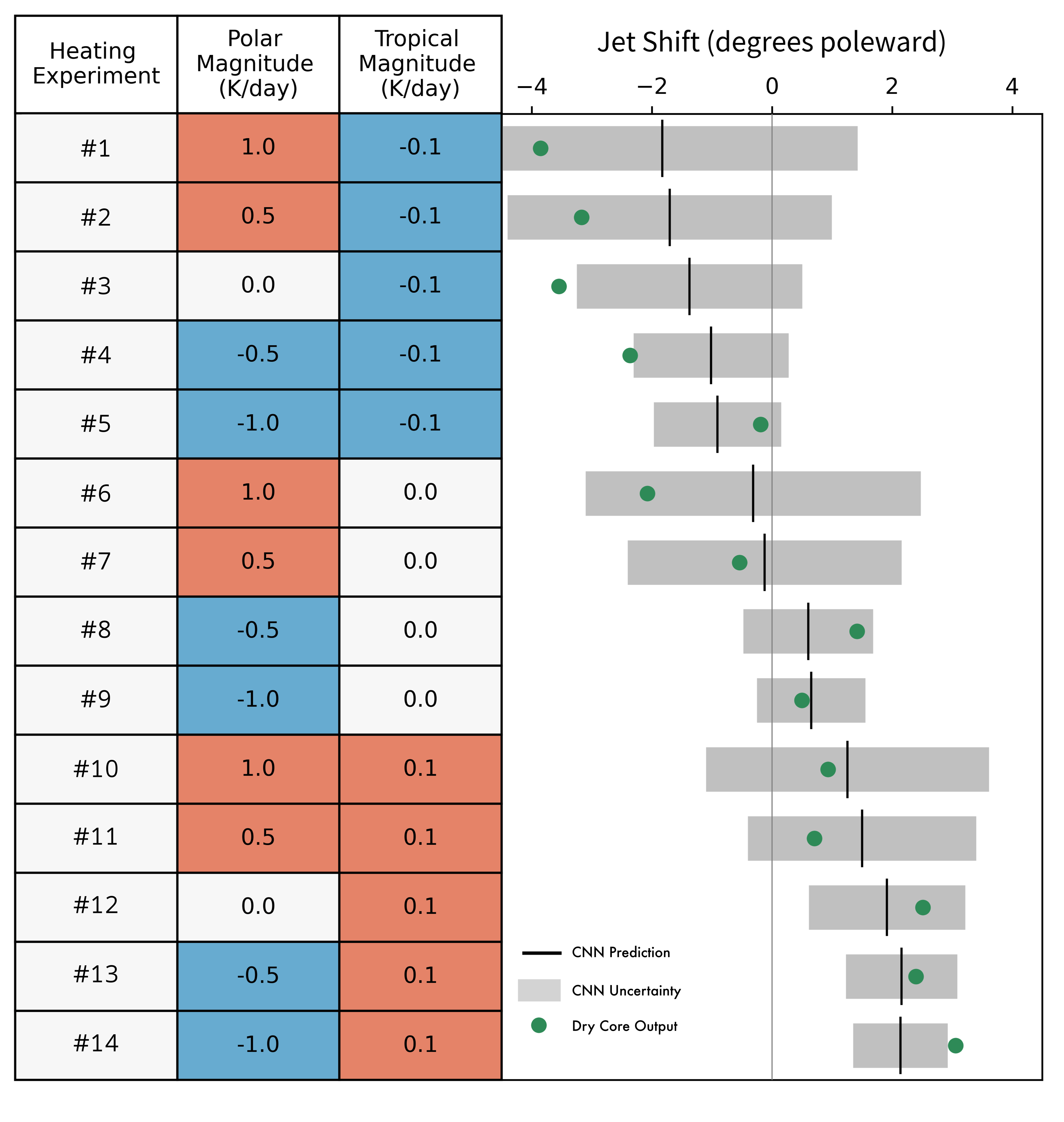}}
  \caption{The 14 heating experiments, their resulting jet shift in the dry core, and the jet shift as predicted by the CNN. The left side includes the magnitude of each Gaussian shaped thermal forcing. The right side shows the true forced dry core jet shift (green dots), the predicted CNN jet shift (black line), and the CNN-predicted uncertainty (gray boxes; $\pm2\sigma$).}\label{DDCCNN}
\end{figure}

We perform 14 additional forced heating experiments with the dry dynamical core (see Methods and Table A1 experiments \#1-14). These 14 heating experiments are motivated by the tug-of-war on the jet resulting from anthropogenic climate change \citep{Harvey2015, Chen2020, Stendel2021}. To mimic the tug-of-war, each experiment contains a thermal forcing in the tropical upper troposphere and at the polar surface. The left side of Figure \ref{DDCCNN} includes the magnitude of each Gaussian shaped thermal forcing. The forced jet response from the dry dynamical core experiments and the CNN-predicted jet response from 14 heating experiments are shown on the right side of Figure \ref{DDCCNN}. Comparing the true forced jet shift simulated by the dry core (green dots) and the predicted jet shift by the CNN trained on internal variability (black lines), we see that across all experiments, the CNN accurately captures the sign of the jet shift. Experiments \#1 through \#7 exhibit a negative jet shift and \#8 through \#14 exhibit a positive jet shift. Furthermore, nearly all of the experiments (excluding \#3, \#4, and \#14) have forced jet shifts that fall well within the uncertainty bounds predicted by the CNN ($\pm2\sigma$; gray boxes).  Additional discussion of other heating experiments that do not focus on the tug-of-war on the jet can be found in Appendix C.

Heating experiments \#6, \#7, \#8, and \#9 contain only a thermal forcing at the polar surface (no thermal forcing in the tropical upper troposphere) and are therefore useful for investigating the difference in CNN-predicted uncertainty between polar warming and polar cooling. In heating experiments \#6 and \#7, which contain polar warming, the CNN is less certain (larger $\sigma$), in contrast to heating experiments \#8 and \#9, which contains polar cooling, where the CNN is more certain (smaller $\sigma$).

The CNN's uncertainty when there is a thermal forcing in the upper troposphere of the tropics is more difficult to discern from Figure \ref{DDCCNN} because the CNN's uncertainty is impacted considerably by the thermal forcing at the polar surface. However, heating experiments \#3 and \#12 include only a thermal forcing in the tropical upper troposphere, one cooling and one warming respectively. These heating experiments suggest that the CNN is more certain when the upper tropical troposphere is cooling rather than warming.

\section{Implications for Sensitivity Analysis}\label{DISCUSSION}

Given the CNN’s ability to replicate the sign of the jet stream’s forced response as validated with the additional forced dry dynamical core experiments, we propose that our approach can be deployed as a computationally efficient tool to aid in the design of forced model experiments. To demonstrate this, we next revisit a historical study (Butler et al. 2010; afterward B10) and show how the pre-trained CNN can be used to replicate the study’s results as well as document possible sensitivities not included in the initial work. In B10, differences in dry core atmospheric circulations due to variations in the location and shape of thermal forcings were identified and documented. To test the atmospheric sensitivity, B10 ran multiple experiments with different imposed thermal forcing patterns in the Colorado State University general circulation model \citep{Ringler2000}. Although B10 did not quantify a shift of the jet stream the same way as we do here, the study showed the zonal-mean zonal wind response and discussed which heating experiments resulted in a stronger wind response. From this information, we are able to infer the relative magnitude of the jet shift for each experiment. 

\begin{figure}[h]
 \centerline{\includegraphics[width=40pc]{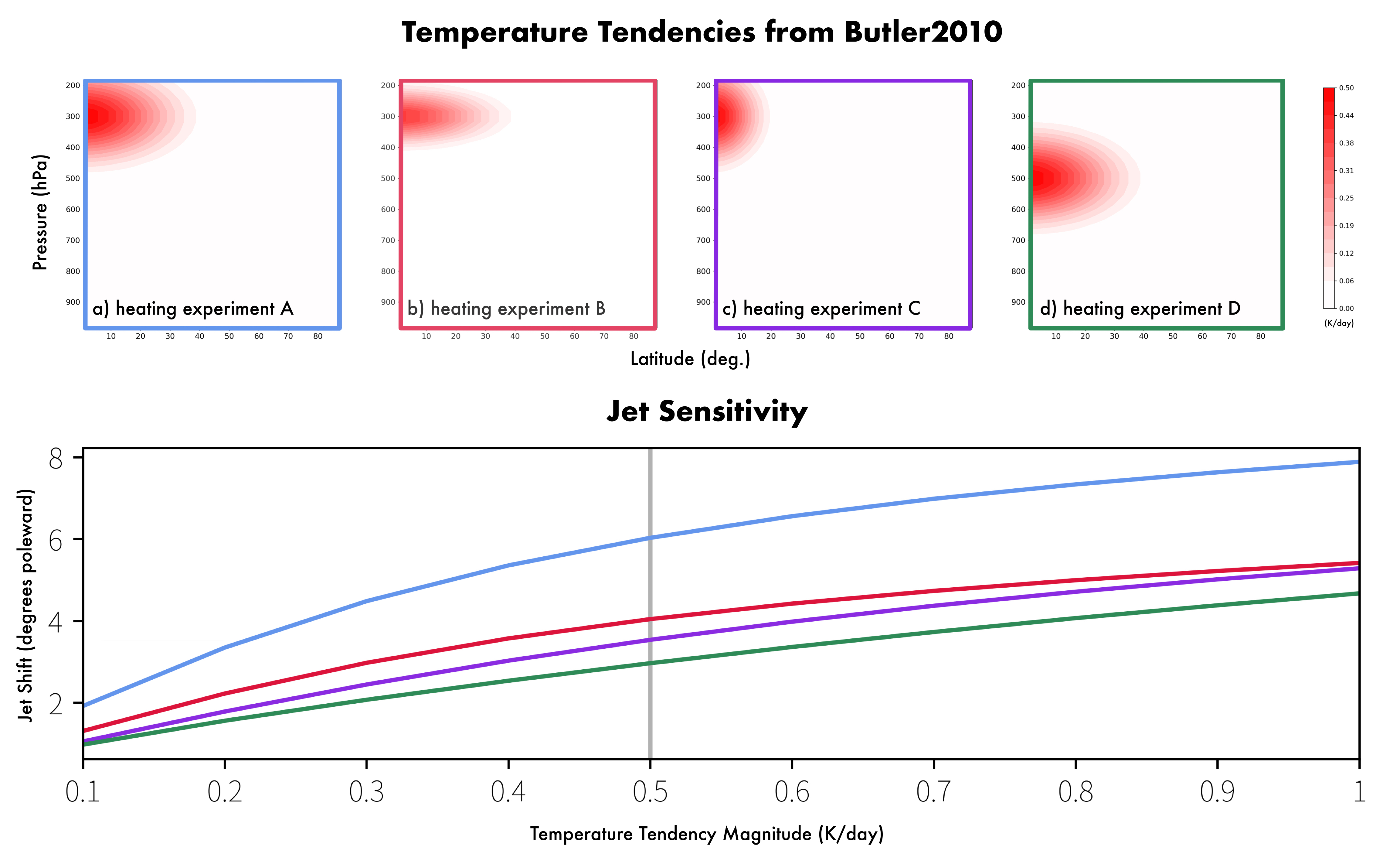}}
  \caption{Panels (a) - (d)  shows example thermal forcings of magnitude 0.5 K day$^{-1}$ representing the four heating experiments used in Butler et al. 2010, Gaussian parameters are found in Table A1 \#26-29. The panel outline corresponds to the predicted jet shifts (y-axis) in panel (e) which are shown as a function of the magnitude of the temperate forcing (x-axis). The vertical gray line at 0.5 K day$^{-1}$ corresponds to the magnitude of the heating experiments used in Butler et al. 2010.}\label{butler2010_plot}
\end{figure}

Here, we focus on four specific heating experiments within B10 that aim to investigate how sensitive the circulation response is to the height and shape of tropical upper tropospheric heating. Examples of the thermal forcing patterns imposed in B10 are shown in Figure \ref{butler2010_plot}a-d and will be referred to as heating experiments \#26, \#27, \#28, and \#29 for this discussion. In the original study, B10 found that the jet shifts poleward in response to all heating experiments, but that the magnitudes of the shifts varied. B10 shows that heating experiment \#26 had the strongest jet response and compressing the forcing vertically (heating experiment \#27) or compressing the forcing in the meridional direction (heating experiment \#28) weakened the wind response. In the last experiment (heating experiment \#29), B10 showed that when the forcing was compressed vertically and moved lower in the troposphere, the wind response was even weaker.

Following the same four heating experiments of B10, we recreated the thermal forcing patterns (see experiments \#26-29 in Table A1) and input them into the CNN with the initial jet location set to the average jet location of the training data (42.4\textdegree). In addition to the magnitude of heating used in B10 (q$_o$ = 0.5 K day $^{-1}$), here the jet sensitivity to the magnitude of the thermal forcing is also included since it is trivial to explore once the CNN is trained. Figure \ref{butler2010_plot}e shows the predicted jet shift from the four heating experiments with varying magnitudes, and the vertical gray line indicates the 0.5 K day $^{-1}$ magnitude used in B10. The CNN predicts the same relative relationship between the heating experiments as found in B10, where heating experiment \#26 exhibits the strongest jet response and heating experiment \#29 exhibits the weakest. Furthermore, by exploring the jet sensitivity to the magnitude of heating, Figure \ref{butler2010_plot}e shows new information about jet sensitivity. For example, as the magnitude of the thermal forcing increases, heating experiment \#27 (compressed vertically) and heating experiment \#28 (compressed meridionally) converge to the same jet shift. Alternatively, when the magnitude of the thermal forcing decreases, heating experiment \#28 and heating experiment \#29 (compressed vertically and lower in the troposphere) converge to the same jet shift. These two results could be confirmed with a few targeted forced dry core simulations (not done here), though it was the ability provided by the CNN to quickly explore jet shifts in response to thermal forcings that allowed us to discover these possible jet sensitivities.

In this chapter, we show a successful example of using the CNN to explore the jet sensitivities inside the dry dynamical core. The CNN is not perfect in its predictions which may be due to a lack of predictability, non-optimal training of the CNN, a breakdown of the FDT, or a combination of the three. However, as demonstrated throughout this paper, the comparisons in the CNN-predicted jet shifts have the same sign and similar magnitudes to the forced jet shifts from the dry core in response to a range of temperature tendencies and once trained, can make predictions quickly. We wish to emphasize that this method should not replace the need to run designed climate model experiments. Rather, training CNN on an existing long control run could provide the opportunity to explore a large number of forcing experiments before any forced model runs are simulated and be especially helpful for planning forced experiments in dynamic model simulations.

\section{Conclusions}\label{CONCLUSION}

We explore the jet stream's response to external forcing by training a CNN on smoothed temperature tendencies from a dry dynamical core long control run to predict a shift in the jet stream’s location 30 days later. The main motivation of this work is to explore the potential for training a CNN on internal variability alone and then using it to examine possible nonlinear responses of the jet to tropospheric thermal forcing that more closely resemble anthropogenic climate change. Since the CNN is trained entirely on data from a control simulation, it exclusively learns from internal variability. Nevertheless, by comparing the CNN-predicted jet shifts to established baselines, peer reviewed literature, and additional dry core heating experiments, we show that the CNN can predict the forced jet shift to sustained forcing. The trained CNN is then used to investigate jet sensitivities to scenarios that mimic the tug-of-war between the tropics and poles under anthropogenic climate change. Given the CNN's ability to predict the jet response to thermal forcings, we propose training a CNN on long control runs that are increasingly becoming more available to explore model sensitivities to various forcings as a tool to aid in early stage climate model experiment design. Future work could include extending this method to evaluate whether it generalizes to different experimental frameworks, including, but not limited to, evaluating it with three-dimensional predictors, training on coupled climate model simulations, and learning complex non-linear climate responses from forced simulations where FDT does not hold.

\section*{Acknowledgments}
CC and EAB are supported, in part, by NSF CAREER AGS-1749261 under the Climate and Large-scale Dynamics program. The authors thank Marybeth Arcodia for providing valuable support during the writing process.

\section*{Data availability}
The code used to prepare the data can be found on github \url{https://github.com/connollyc152/DDC_jet_sensitivity} and will be assigned a permanent doi on Zenodo upon publication. Data is available upon request and will be made available on Zenodo and given a doi upon publication.

\newpage{}
\section*{Appendix A}
\section*{Table of two-dimensional Gaussians parameters}

\begin{table}[h!]
 \centerline{\includegraphics[width=28pc]{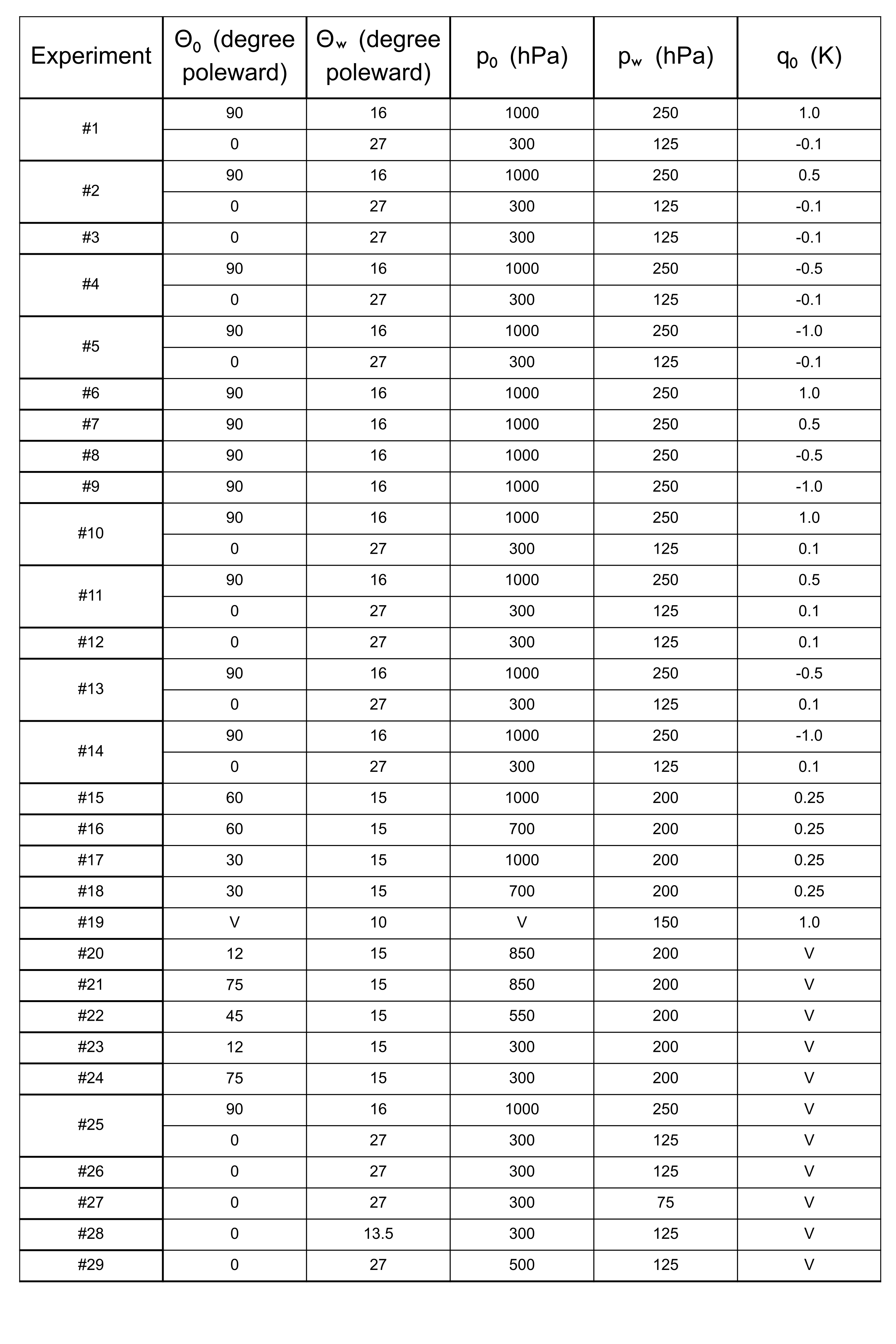}}
  \caption{Parameters for two-dimensional Gaussians for the forced dry core heating experiments and CNN thermal forcings, "V" indicates varying parameter.}\label{HG}
\end{table}

\newpage{} 

\bibliography{ArxivTemplate} 
 
\end{document}